\documentstyle[12pt]{article}
\begin{document}
\baselineskip 1cm
\begin{center}
{\bf 5D Schwarzschild-like Spacetimes with Arbitrary Magnetic Field}\\
by\\
Tonatiuh Matos\footnote[1]{Work partialy supported by CONACyT-M\'exico}\\
Centro de Investigaci\'on y de Estudios Avanzados del I.P.N.\\
Departamento de F\'\i sica\\
Apdo. Postal 14-740, C.P. 07000\\
M\'exico, D.F.
\end{center}
\vskip 1cm
\noindent
{\bf Abstract}.  We find a new class of exact solutions of the
five-dimensional Einstein equations whose corresponding four-dimensional
spacetime possesses a Schwarzschild-like behavior.  The
electromagnetic potential depends on a harmonic function and can be choosen
to be of a monopole, dipole, etc. field.  The solutions are
asymptotically flat and for vanishing magnetic field the four metrics are
of the Schwarzschild solution. The spacetime is singular in $r=2m$
for higher multipole moments, but regular for monopoles or vanishing
magnetic fields in this point.
The scalar field posseses a singular behavior.
\vfil\eject
\indent

In the last years there has been a great interest in the study of exact
solutions of actions of type
\begin{equation}
S=\int d^4x \sqrt{-g}[-R + 2(\nabla\Phi)^2 + e^{-2\alpha\Phi}F^2],
\end{equation}
\noindent
because it reduces to the four-dimensional low energy Lagrangian for string
theory for $\alpha=1$, to the Einstein-Maxwell-Scalar theory for $\alpha=0$
and it also reduces to five-dimensional gravity, for $\alpha=\sqrt{3},$
after dimensional reduction. Some exact solutions of the field equations
for charged bodies of this action are known \cite{horne}. It seems that
the properties of electrically charged solutions depend on the value of
$\alpha$, but they are only different for the extrem case $\alpha=0$
\cite{garfinkle}.
In this latter we want to show that magnetic fields donot alter the properties
of the spacetime for static bodies, at less for $\alpha=\sqrt{3}$. We
present a set of exact static solutions of this Lagrangian where the
magnetic field depend on a harmonic map, and can be chosen to be of a
monopole, dipole, quadripole, etc.

Einstein theory is a good model for descrybing gravitational interactions in
the universe. Nevertheless there are some phenomena in cosmos where
gravitation is interacting with electromagnetism. Such is the case for
example, in planets and stars possesing a magnetic field as the earth or
the sun. Our galaxy posseses also a magnetic field and there is not yet a
convincening explanation for it. One would wait that Einstein-Maxwell theory
should give such an explanation by means of a simple exact solution possesing
a magnetic field like the celestial bodies. There is an exact solution of
the Einstein-Maxwell equations containing magnetic dipole moment satisfying
the required stationary and static limits \cite{manko}.
But it is not simple at all.
5-dimensional (5D) theory is an alternative model for understanding
gravitational and electromagnetical interactions together.
In this work we want to show that there exist a class of very simple
exact solutions of the 5D Einstein equations possesing magnetic fields
which 4-dimensional metric behaves like the Schwarzschild solution.
In a past work \cite{matos}
we developed a method for generating exact solutions of the
five-dimensional Einstein equations with a $G_3$ group of motion,
puting the solutions in terms of two harmonic maps $\lambda$ and $\tau$.
These solutions can be also interpreted as solutions of Lagrangian
(1) for the case $\alpha=\sqrt3$.
We separated the solutions in five tables
(tables III-VII) for the one- and two-
dimensional (abelian and nonabelian) subgroups of SL(2,R),
in the spacetime and the
potential space, and demostrated that many of the well-known solutions are
contained in these tables. In this letter we want to present
a set of new solutions which belong to the class of solutions $i,j$
and $k$ of table VI in reference \cite{matos},
specialising the harmonic maps,
because it represents a class of very good behaved solutions,
if we choose the harmonic
maps $\lambda$ and $\tau$ conveniently.  In terms of the five potentials
\cite{neuge} the solutions are\\
\[
i)\;\;\; \chi =\frac{a_1e^{q\tau}+a_2 e^{-q\tau}}{g_{22}} \;\;\;\;\;
g_{22}=be^{q\tau}+ce^{-q\tau} \;\;\;\; b+c=\frac{1}{I_0}
\]
\[
j)\;\;\; \chi =\frac{a_1\tau +a_2}{g_{22}} \;\;\;\;\;
g_{22}=b\tau +\frac{1}{I_0}\;\;\;\;\;\;\;\;\;\;
\]
\[
k)\;\;\; \chi =\frac{a_1e^{iq\tau} +{\bar a}_1e^{-iq\tau}}{g_{22}}
\;\;\;\;\;
g_{22}=be^{iq\tau}+{\bar b}e^{-iq\tau} \;\;\;\; b+{\bar b}=\frac{1}{I_0}
\]
\noindent
(here we have set $\alpha +\beta =0$ in table VI of ref. \cite{matos}).  The
gravitational potential and the scalar potential are the same for all cases
given by\\
\[
f=\frac{e^{\wedge\lambda}}{\sqrt{I_0 g_{22}}} \;\;\;\; , \;\;\;
I^2=\frac{I_0{\bar e}^{\frac{2}{3}\wedge\lambda}}{g_{22}} \;\;\; ,
\]
\noindent
where $a_1 , a_2, q, b, c, I_0$ and $\wedge$ are constants restricted by
$bcq^{2}=I_0 \delta \neq 0,$ while the electrostatic and rotational potentials
vanish, i.e. $\psi =\epsilon =0$.  Now it is easy to write the spacetime
metric.  Let us write it in Boyer-Lindquist coordinates\\
\[
\rho =\sqrt{r^2+2mr}sin\theta \;\;\; , \;\;\; z =(r-m)cos\theta .
\]
\noindent
In this coordinates the five-metric reads\\
\[
dS^2 =\frac{1}{I}\left\{ \frac{1}{f}e^{2k}\left[1-\frac{2m}{r}+
\frac{m^2\sin^2\theta}{r^2}\right]
\left[\frac{dr^2}{1-\frac{2m}{r}}+r^2 d\theta^2\right]
\right.\]
\[
\left.+\frac{1}{f}\left(1-\frac{2m}{r}\right)r^2 sin^{2}\theta d\varphi^2
-fdt^2\right\} +
I^2(A_3d\varphi +dx^5)^2
\]
\noindent
The expression in brackets $\{...\}$, is intepreted as the four-dimensional
metric in the five-dimansional theory and corresponds to the spacetime
metric of Lagrangian (1).
The functions $k$ and $A_3$ are completly determined by the potentials
$f, \chi$ and $\kappa^2 = I^3$\\
\[
k_{,\zeta} =\rho [\frac{(f_{,\zeta})^2}{2f^2}+\frac{1}{2f}
\frac{(\chi _{,\zeta} )^2}{\kappa^2}+\frac{2}{3}
\frac{(\kappa _{,\zeta} )^2}{\kappa^2}]=\frac{\rho}{2}
[\frac{4}{3}\wedge^2(\lambda _{,\zeta})^2 +q^2 (\tau _{,\zeta})^2]
\]
\[
A_{3,\zeta}=-\frac{\rho}{f\kappa^2}\chi_{,\zeta}=-\rho\tau_{,\zeta}
\]
\[
A_{3,\bar\zeta}=\rho\tau_{,\bar\zeta} \;\;;\;\;\;\;\;\; \zeta =\rho + iz
\]
\noindent
Observe that the function $A_3$ is integrable because $\tau$ fulfills the
Laplace equation $(\rho\tau_{,\zeta})_{,\bar\zeta}+
(\rho\tau_{,\bar\zeta})_{,\zeta}=0$.  In reference \cite{becerril} a set of
solutions of
the Laplace equation and their corresponding magnetic potential $A_3$ is
listed.  Two examples are\\
\[
a) \;\;\; \tau =\tau_o {\em ln} (1-\frac{2m}{r}) \;\;\;\;
A_{3}=2\tau_0 m(1-cos \theta )
\]
\[
b) \;\;\; \tau =\frac{\tau_o m^2cos\theta}{(r-m)^2-m^2cos^2\theta}\;\;\;\;
A_3=\frac{m^2\tau_o(r-m)sin^2\theta}{(r-m)^2-m^2cos^2\theta}
\]
\noindent
written in Boyer-Lindquist coordinates.  The magnetic potential $a)$ and
$b)$ represents a magnetic monopole and a magnetic dipole, respectively.  In
general the harmonic function $\tau$ determines the magnetic field in the
solution and can be choosen to obtain monopoles, dipoles, quadripoles, etc.
fields.  The harmonic function $\lambda$ determines the gravitational
potential $f$.  Let us choose $\lambda =\lambda_0{\em ln}
(1-\frac{2m}{r})$.  The five metric transforms to\\
\[
dS^2=\sqrt{\frac{{\em g}_{22}}{I_0}}
\left(1-\frac{2m}{r}\right)^{\frac{1}{3}\wedge\lambda_0}
\left\{ \frac{(1-\frac{2m}{r}+
\frac{m^2sin^2\theta}{r^2})^{1-\frac{4}{3}\wedge^2 \lambda^{2}_0}}
{(1-\frac{2m}{r})^{\wedge\lambda_0-\frac{4}{3}\wedge^2 \lambda^2_0}}
\sqrt{I_0{\em g}_{22}}e^{2k_1}
\left(\frac{dr^2}{1-\frac{2m}{r}}+r^2d\theta^2\right)
\right.\]
\[
\left.+\left(1-\frac{2m}{r}\right)^{1-\wedge\lambda_0}
\sqrt{I_0{\em g}_{22}}r^2sin^2\theta d\phi^2
-\frac{1}{\sqrt{I_0{\em g}_{22}}}
\left(1-\frac{2m}{r}\right)^{\wedge\lambda_0} dt^2\right\}
\]
\begin{equation}
+I^2(A_3d\phi +dx^5)^2
\end{equation}
\noindent
where $k_{1_{,\zeta}}=\frac{1}{2}q^2\tau_0(\tau_{,\zeta})^2 , {\em g}_{22}$
and $A_{3}$ are determined only by the harmonic function $\tau$.  If we choose
$\tau$ to vanish for some limit $r>>m,$ (the two examples $a)$
and $b)$ fulfill this
condition) then the metric (2) is asymptotically flat.  If $\tau$ and $m$
vanish, metric (2) is flat.\\
\indent
If we put $\wedge =-2,\lambda_0 =-\frac{1}{2}$ in (2) we can interprete $m$
as the mass para\-meter and $\sqrt{I_0{\em g}_{22}}$ as the contribution
of the magnetic field to the metric.  In this case, metric (2) reads \\
\[
dS^2=\frac{1}{I}\left\{ \sqrt{I_0{\em g}_{22}}e^{2k_2}
(\frac{dr^2}{1-\frac{2m}{r}}+
r^2d\theta^2)
\right.\]
\[
\left.+\sqrt{I_0{\em g}_{22}}r^2sin^2\theta d\phi^2-\frac{1}{\sqrt{I_0{\em
g}_{22}}}
\left(1-\frac{2m}{r}\right)dt^2\right\}+I^2(A_3d\phi +dx^5)^2
\]
\begin{equation}
I^2=\frac{I_0(1-\frac{2m}{r})^{-\frac{2}{3}}}{{\em g}_{22}}
\end{equation}
\indent
This metric can be interpreted as a magnetized Schwarzschild solution in
five-dimensional gravity.  The difference to a previous one \cite{matos1}
is that in
metric (3) the magnetic potential can be choosen in many ways.  If the
magnetic field $A_3$ in (3) vanishes, the expression in brackets \{...\} is
just the Schwarzchild metric.  Therefore we can interpret $r=2m$ as the
horizon of the four metric. Observe that the presens of the magnetic field
doesnot alter the horizon of the metric, conserving the main feature of
its topology. Nevertheless the scalar field do.
We can see that the scalar potential
tends very fastly to $I_0$ for $r>>2 m$ and is singular for $r=2M$.
If we interpret the expression in brackets $\{...\}$ as the spacetime
metric, we find that its Rimannian invariant $R^{abcd}R_{abcd}$ and
its Ricci invariant $R^{ab}R_{ab}$ are singular for $r=2M$,(but not
its scalar curvature $R$), when $\tau$ depends on $\theta$.
This is so for the case when $A_3$ represents the magnetic field
of a dipole, but when $A_3$ represents a monopole, all invariants
remain regular on $r=2M$. So, one expects that $r=2M$ is a coordinate
singularity when $A_3$ is a monopole field, but the spacetime is really
singular for higher multipole moments in this point, and are not
black holes. However for geodesical
trajectories arround the surface $r>2M$, the effective potential
is regular for $r=2M$ even for magnetic dipole fields, but the scalar
field increases without bound for all these cases when $r$ aproaches
$2M$. The scalar field I is topologically the radius of the
fifth-dimension which is a circle. This circle has constant radius
for $r>>2M$, but  tends to a line when  $r$ aproaches $2M$.
That means that the scalar potential is really important only near
of the horizon but desapears very fastly far away of it.
One would suspect that the properties of the geometry change
near the horizon with respect to Schwarzschild's geometry due to
the interaction of the scalar field. That means that the relevant
modifications of the Schwarzshcild's geometry is not due to the
magnetic field but due to the scalar interaction. The geodesic
motion in this spacetime will be publish elswhere \cite{nora}.
\\



\begin{thebibliography}{9}
\bibitem{horne} G. W. Gibbons and K. Maeda. {\em Nucl. Phys. {\bf B298}},
(1988), 2677
J. N. Horne and G. T. Horowitz {\em Phys. Rev. {\bf
D46}}, (1992), 1340.
\bibitem{garfinkle}D. Garfinkle, G. T. Horowitz and A. Strominger.
{\em Phys. Rev. {\bf D43}}, (1991), 3140.
A. Shapere, S Trivesi and  F. Wilczek. {\em Mod. Phys. Lett. {\bf A6}},
(1991), 2677.
\bibitem{manko} V. S. Manko and N. R. Sibgatullin {\em Phy. Rev. {\bf
D46}}, (1992), R4122.
\bibitem{matos} T. Matos.{\em J.Math.Phys.} to appear.
\bibitem{neuge} G. Neugebauer {\em Habilitationsschrift} (1969) (Doctor in
Science Thesis)-Jena.
\bibitem{becerril} R. Becerril and T. Matos. {\em Phys. Rev. {\bf D46}},
(1992), 1540.
\bibitem{matos1} T. Matos {\em Phys. Rev. {\bf D38}}, (1988), 3008.
\bibitem{nora} T. Matos and N. Breton. To be published.
\end{thebibliography}
\end{document}